\documentclass[aps,preprint,nofootinbib]{revtex4}
\usepackage{amssymb}
\usepackage{amsmath}
\usepackage{bm}
\begin{document}
\title{Inflationary dark energy from a condensate of spinors in a 5D vacuum}
\author{$^{1, 2}$ Pablo Alejandro S\'anchez\footnote{pabsan@mdp.edu.ar}, and $^{1, 2}$Mauricio
Bellini\footnote{mbellini@mdp.edu.ar}} \address{$^{1}$
Departamento de F\'{\i}sica, Facultad de Ciencias Exactas y
Naturales, \\
Universidad Nacional de Mar del Plata, Funes 3350, (7600) Mar del
Plata, Argentina. \\ \\
$^{2}$ Instituto de Investigaciones F\'{\i}sicas de Mar del Plata
(IFIMAR), Consejo Nacional de Investigaciones Cient\'{\i}ficas y
T\'ecnicas (CONICET), Argentina.}
\begin{abstract}
What is the physical origin of dark energy? Could this energy be
originated by other fields than the inflaton? In this work we
explore the possibility that the expansion of the universe can be
driven by a condensate of spinors. These spinors are free of
interactions on 5D relativistic vacuum in an extended de Sitter
spacetime. The extra coordinate is considered as noncompact. After
making a static foliation on the extra coordinate, we obtain an
effective 4D (inflationary) de Sitter expansion which describes an
inflationary universe. In view of our results we conclude that the
condensate of spinors here studied could be an interesting
candidate to explain the presence of dark energy in the early
universe. \vskip 1cm \noindent {\bf Essay awarded by the Gravity
Research Foundation with the Honorable Mention - Annual Essays
Competition 2013.}
\end{abstract}
\maketitle

Modern versions of 5D General Relativity abandon the cylinder and
compactification conditions used in original Kaluza-Klein (KK)
theories, which caused problems with the cosmological constant and
the masses of particles, and consider a large extra dimension. In
particular, the Induced Matter Theory (IMT)\cite{IMT} is based on
the assumption that ordinary matter and physical fields that we
can observe in our 4D universe can be geometrically induced from a
5D Ricci-flat metric with a space-like noncompact extra dimension
on which we define a physical vacuum. From the mathematical point
of view, the Campbell-Magaard Theorem (CMT)\cite{campbell} serves
as a ladder to go between manifolds whose dimensionality differs
by one. This theorem implies that every solution of the 4D
Einstein equations with arbitrary energy momentum tensor can be
embedded, at least locally, in a solution of the 5D Einstein field
equations in a relativistic vacuum: $G_{AB}=0$\footnote{We shall
consider that letters $A,B$ run from $0$ to $4$ and Greek letters
rum from $0$ to $3$.}. Because of this, the stress-energy may be a
4D manifestation of the embedding geometry. Therefore, by making a
static foliation on the space-like extra coordinate of an extended
5D de Sitter spacetime, it is possible to obtain an effective 4D
universe that suffered an exponential accelerated expansion driven
by an effective scalar field with an equation of state typically
dominated by vacuum\cite{ua,ua1,LB,B}. The most conservative
assumption is that the energy density $\rho =P/\omega $ is due to
a cosmological parameter which is constant and the equation of
state is given by a constant $\omega =-1$, describing a vacuum
dominated universe with negative pressure $P$ and energy density
$\rho$. This is the simplest version of inflation in the IMT.
However, this is not the unique.

There is an intriguing problem in modern cosmology that relies to
explain the physical origin of the cosmological constant, which is
responsible for the exponential expansion of the early universe,
as well as the present day accelerated expansion of the universe.
The standard explanation for the early universe expansion is that
it is driven by the inflaton field. Many cosmologists mean that
such acceleration (as well as the present day accelerated
expansion of the universe) could be driven by some exotic energy
called {\em dark energy}. But what is the physical origin of dark
energy? Could this energy be originated by other fields than a
scalar field? Most versions of inflationary cosmology require of
one scalar inflaton field which drives the accelerated expansion
of the early universe with an equation of state governed by the
vacuum\cite{1}. The parameters of this scalar field must be rather
finely tuned in order to allow adequate inflation and an
acceptable magnitude for density perturbations. The need for this
field is one of the less satisfactory features of inflationary
models. Consequently, we believe that it is of interest to explore
variations of inflation in which the role of the scalar field is
played by some other field\cite{2,3}. Recently has been explored
the possibility that such expansion can be explained by a
condensate of dark spinors\cite{Lee}. This interesting idea was
recently revived in the framework of the Induced Matter Theory
(IMT). In this work we revise this idea, but from a five
dimensional vacuum.

As 5D vacuum we understand a purely kinetic 5D lagrangian for
fields minimally coupled to gravity which are in a 5D perfect
fluid on a, at least, 5D Ricci-flat spacetime. In particular, we
shall consider a 5D Riemann-flat spacetime
\begin{equation}\label{met1}
dS^{2}=\left(\frac{\psi}{\psi_{0}}\right)^{2}\left[dt^{2}-e^{\frac{2t}{\psi_{0}}}(dx^{2}+dy^{2}+dz^{2})\right]-d\psi^{2},
\end{equation}
where ${t, x, y, z}$ are the usual local spacetime coordinate
system and $\psi$ is the noncompact space-like extra dimension. If
we foliate $\psi=\psi_{0}=H^{-1}_0$, the resulting 4D hypersurface
describes a de Sitter spacetime. This means that a relativistic
observer moving with the penta-velocity $U^{\psi}=0$, will be on a
spacetime which has a scalar curvature $^{(4)}
R=12/\psi^2_0=12\,H^2_0$, such that the constant Hubble parameter
(and hence the cosmological constant: $\Lambda = {3 H^2_0\over
8\pi G}$) is determined  by the foliation from the geometrical
point of view. So, after making the foliation, we get the
effective 4D metric
    \begin{equation}\label{met2}
        dS^{2}\rightarrow ds^{2}=dt^{2}-e^{2H_{0}t}dr^{2},
    \end{equation}
which describes a 3D spatially flat, isotropic and homogeneous de
Sitter expanding universe with a constant Hubble parameter
$H_{0}$.

From the physical point of view we shall consider the action
${\cal S} = \int d^5 x\, \left[ {{\cal{ R}}\over 16 \pi G} +
\mathcal{L}_{eff}\right]$, where the 5D Lagrangian density for
free massless spinors is $
\mathcal{L}_{eff}=-\frac{1}{2}(\nabla_{A}\overline{\Psi})(\nabla^{A}\Psi)$.
The equation of motion for the spinor $\Psi$ takes the form [The
same procedure yields an identical equation for the field
$\overline{\Psi}$ .]
    \begin{equation}\label{mot}
        g^{AB}\nabla_{A}\nabla_{B}\Psi=0.
    \end{equation}

A clever transformation on the spinor components is $   \Psi=
        \left(
            \begin{array}{c}
                \varphi_{1} \\
                \varphi_{2} \\
        \end{array}
        \right)$, where components are $ \varphi_{1}=
        \left(
          \begin{array}{c}
            \psi_{1} \\
            \psi_{2} \\
          \end{array}
        \right)
        {\rm and} \enskip
        \varphi_{2}=
        \left(
          \begin{array}{c}
            \psi_{3} \\
            \psi_{4} \\
          \end{array}
        \right)$,
can be followed by the conformal mapping
$\Phi_{+}=\varphi_{1}+i\varphi_{2}$ and
$\Phi_{-}=\varphi_{1}-i\varphi_{2}$. Rewriting the equations of motion
in terms of the new fields, we
get decouple the equation for $\Phi_{+}$, while the other coupling becomes
a source for the equation for $\Phi_{-}$\footnote{Here, we have adopted the following convention[$\overrightarrow{\sigma}$ denotes the Pauli vector]:
    \begin{eqnarray}
        \widehat{{\O}}\varphi&=&\left(\frac{\psi_{0}}{\psi}\right)^{2}\frac{\partial^{2}\varphi}{\partial t^{2}}-\left(\frac{\psi_{0}}{\psi}\right)^{2}
        e^{-\frac{2t}{\psi_{0}}}\ \nabla^{2}\varphi-\frac{\partial^{2}\varphi}{\partial \psi^{2}}.\nonumber
           \end{eqnarray}}
\begin{eqnarray}
\widehat{{\O}}\Phi_{+}+\frac{2\psi_{0}}{\psi^{2}}\
\frac{\partial\Phi_{+}}{\partial t}-\frac{4}{\psi}\
\frac{\partial\Phi_{+}}{\partial
\psi}-\frac{5}{4\psi^{2}}\Phi_{+}&=&0, \label{a}
        \\
\widehat{{\O}}\Phi_{-}+\frac{4\psi_{0}}{\psi^{2}}\
\frac{\partial\Phi_{-}}{\partial t}-\frac{4}{\psi}\
\frac{\partial\Phi_{-}}{\partial
\psi}+\frac{7}{4\psi^{2}}\Phi_{-}&=&\frac{i2\psi_{0}}{\psi^{2}}\
e^{-\frac{t}{\psi_{0}}}\
\overrightarrow{\sigma}\cdot\overrightarrow{\nabla}\Phi_{+}.
\label{b}
\end{eqnarray}
Notice that the equation of motion (\ref{a}) for $\Phi_{+}$ is
homogeneous, but the equation (\ref{b}) for $\Phi_{-}$ has a
source where $\vec{\nabla}\Phi_{+}$ appears coupled to
$\vec{\sigma}$. The equation (\ref{a}) can be factored as a
product of functions $\Phi_{+} \propto \Lambda_1(\psi)
\,e^{i\vec{k}.\vec{r}}\, \xi_k(t)$ and can be proved that for
cosmological scales [i.e., for $k \ll H_0 \, e^{H_0 t}$], the
equation (\ref{b}) also can be factored as $\Phi_{-} \propto
\Lambda_2(\psi) \,e^{i\vec{k}.\vec{r}}\, \chi_k(t)$\footnote{The
solutions on cosmological scales for the relevant functions
$\Lambda_1(\psi)$, $\Lambda_2(\psi)$, $\xi_k(t)$ and $\chi_k(t)$
are [$M^2_1= {(n-3/2)^2-1\over \psi^2_0}$ and $[M_2(m,n)]^2 =
{(m-3/2)^2-(n-3/2)^2+4\sqrt{(n-3/2)^2+3}-10\over \psi^2_0}$ are
separation constants that only can take certain values: $n\geq 3$
with $m> 3/2+\sqrt{3+\left(\sqrt{(n-3/2)^2+3}-2\right)^2}$]
\begin{eqnarray} &&  \Lambda_1(\psi) = \lambda_1 \left(\psi/\psi_0\right)^{-n}, \quad
\Lambda_2(\psi) = \lambda_2 \left(\psi/\psi_0\right)^{-m}, \quad
\xi_k(t) \simeq  C_1\, e^{\left(\nu -2\right)\,t/\psi_0}\, \left[
(k/2)
\psi_0\right]^{-\nu},\\
&&  \chi_k(t) \simeq C_2 \,
e^{\left(\mu-1\right)\,t/\psi_0}\,\left[(k/2) \psi_0\right]^{-\mu}
+ F(t)
\end{eqnarray}
where $\left.F(t)\right|_{t\rightarrow \infty}\rightarrow  C$,
$\nu=\sqrt{4+M^2_1 \psi^2_0}$ and $\mu= \sqrt{\nu^2-4\nu
+4+M^2_2\psi^2_0}$.}. After few calculations, the Lagrangian
density ${\cal L}_{eff}$, written in term of the new fields
$\Phi_{+}$ and $\Phi_{-}$, results
    \begin{equation}\label{lag5}
        \mathcal{L}_{eff}=-\frac{1}{4}\left(\nabla_{A}\overline{\Phi}_{+}\ \nabla^{A}\Phi_{+}
        +\nabla_{A}\overline{\Phi}_{-}\
        \nabla^{A}\Phi_{-}\right).
    \end{equation}
On the other hand the 5D energy-momentum tensor is represented by
$^{(5)}T_{AB}=2\frac{\delta\mathcal{L}_{eff}}{\delta
g^{AB}}-g_{AB}\mathcal{L}_{eff}$. This procedure take place in a
5D vacuum.\\

\noindent {\bf{\em 4D dynamics}}\\

Now we consider a static foliation on the 5D metric (\ref{met1}).
The resulting 4D hypersurface after making $\psi=\psi_0$ describes
a de Sitter spacetime, which rises on the metric (\ref{met2}).
From the relativistic point of view an observer who moves with the
penta velocity $U^\psi=0$ on a 4D hypersurface that has a scalar
curvature $^{(4)} R=12/\psi^2_0=12\,H^2_0$, such that the Hubble
parameter $H_0$ is defined by the foliation $H_0=\psi_0^{-1}$. He
will undergo a co-moving de Sitter expansion with the universe.

The effective 4D Lagrangian density (\ref{lag5}) is expressed in
terms of the fields $\Phi_{\pm}(x^{\mu},\psi_0)$, which can be
thought of as two minimally coupled bosons
$\mathcal{L}_{eff}=-\frac{1}{4}\left[\nabla_{\mu}\overline{\Phi}_{+}\
\nabla^{\mu}\Phi_{+}+\nabla_{\mu}\overline{\Phi}_{-}\,\nabla^{\mu}\Phi_{-}\right
]+ V\left(\Phi_{+},\Phi_{-}\right)$. Since $\nabla_4 \phi_{+} =
{\partial \phi_{+}\over
\partial\psi}=-(n/\psi)\,\phi_{+}$ and $\nabla_4 \phi_{-} =
{\partial \phi_{-}\over
\partial\psi}=-(m/\psi)\,\phi_{-}$, hence the
effective 4D potential $
V\left(\Phi_{+},\Phi_{-}\right)=-\left.{1\over 4}
\left[\nabla_{4}\overline{\Phi}_{+}\
\nabla^{4}\Phi_{+}+\nabla_{4}\overline{\Phi}_{-}\
\nabla^{4}\Phi_{-}\right]\right|_{\psi=1/H_0}= {H^2_0\over 4}
\left(n^2 \Phi_{+}\overline\Phi_+ + m^2
\Phi_{-}\overline\Phi_{-}\right)$ induced by the static foliation
on the fifth coordinate $\psi=\psi_0=1/H_0$, is the responsible to
provide the dynamics of the fields $\Phi_{\pm}(x^{\mu},\psi_0)$ on
the effective 4D hypersurface, such that $\omega=P/\rho=-1$.
However, after making a constant foliation we shall recover the
same fields in an effective 4D de Sitter (and curved) spacetime.
Some terms coming from the 5D geometry become potentials in the
effective 4D energy-momentum tensor. In turn, these terms become
effective mass of particles in 4D in the equation of motion. The
energy density and pressure related to these fields are obtained
from the diagonal part of the energy-momentum tensor written in a
mixed manner
\begin{small}
\begin{eqnarray}
\rho &=& \left<E\left| \frac{1}{4} \left[\left(\nabla_0
\Phi_+\right)^2 + \left(\nabla_0 \Phi_-\right)^2\right] -
\frac{e^{-2 H_0 t}}{4} \left[ \vec{\nabla}\Phi_{-} .
 \vec{\nabla} \bar\Phi_{-} + \left(\vec{\nabla}\Phi_{+} .\vec{\nabla}\bar{\Phi}_{+}\right)^2\right]
\right. \right.\nonumber \\
& + &\left.\left.\left.
V\left(\Phi_+,\Phi_-\right)+F^0_{\,\,0}\right|E\right>
\right|_{\psi=1/H_0},
\\
 P &=& \left<E\left| \frac{1}{4} \left[\left(\nabla_0
\Phi_+\right)^2 + \left(\nabla_0 \Phi_-\right)^2\right] -
\frac{e^{-2 H_0 t}}{12} \left[ \vec{\nabla}\Phi_{-} .
 \vec{\nabla} \bar\Phi_{-} + \left(\vec{\nabla}\Phi_{+} .\vec{\nabla}\bar{\Phi}_{+}\right)^2\right]
\right.\right. \nonumber \\
& -& \left.\left. \left.V\left(\Phi_+,\Phi_-\right)+F^i_{\,\,j}
\delta^i_j\right|E\right>\right|_{\psi=1/H_0}, \nonumber
\end{eqnarray}
\end{small}
where $F^0_{\,\,\,0} = C_3/\pi \left[{H^7_0\over 8k^2}+{H^5_0\over
k^4}\right]$, $F^i_{\,\,\,j} = C_4/\pi \left[{15 H^7_0\over 32
k^2}+{H^5_0 \over 2 k^4}\right] \delta^i_j$. The stress-energy may
be a 4D manifestation of the embedding geometry. By making a
static foliation on the space-like extra coordinate, it is
possible to obtain an effective 4D universe that suffered an
exponential accelerated expansion driven by the fields
$\Phi_{\pm}$. In order to obtain the equation of state $\omega=-1$
we must require that $\rho=-P=3H^2_0/(8\pi G)$, where
$\left.|E\right>$ is some quantum state such that
$\left<E|V|E\right>$ denotes the expectation value of the
effective 4D potential $V$ on the 3D Euclidean hypersurface.
Because we are interested in the contribution of energy density on
large cosmological scales, the squared gradients in $P$ and $\rho$
can be neglected. On the other hand it is easy to demonstrate that
$\nabla_0 \Phi_{\pm} =
\partial_t \Phi_{\pm} \pm {1\over 4\psi_0} \Phi_{\pm}$. Finally,
we arrive to one possible consistent solution for the energy
density and pressure using the values $\mu=1$ and $\nu=2$,
corresponding to the masses $M_1=0$ and $M_2^2=1/\psi^2_0=H^2_0$
and $n=5/2$ with $m=7/2$. This result corresponds to the choice of
fields which are constants with respect to the time and spectrums
compatible with one scale invariant. The field $\phi_{+}$ is
massless and with spin $1$, but the $\phi_{-}$ is a scalar boson
which could be the responsible for the expansion of the universe
and therefore it could be a good substitute for the inflaton
field, because $\phi_{-}$ can drive the expansion of the universe.
In our picture $\phi_{-}$ is an effective field which became from
a condensate of two entangled spinors that have $2$ components. In
all our analysis we have neglected the role of the inflaton field,
which (in a de Sitter expansion) is freezed in amplitude and
nearly scale invariant. Therefore, the condensate of spinors could
be a interesting candidate to explain the presence of dark energy
in the early universe.

\section*{Acknowledgements}
\noindent The authors acknowledge UNMdP and CONICET Argentina for
financial support.

\end{document}